\documentclass[prb,showpacs,amsmath,amssymb,twocolumn,superscriptaddress]{revtex4-1}
\usepackage{graphicx}
\usepackage{dcolumn}
\usepackage{bm}
\usepackage{amssymb}
\usepackage{amsmath}
\usepackage{epstopdf}
\usepackage{booktabs}
\usepackage{threeparttable}
\usepackage[pdfstartview=FitH, CJKbookmarks=true, bookmarksnumbered=true, bookmarksopen=true, colorlinks, pdfborder=001, linkcolor=blue, anchorcolor=blue, citecolor=blue]{hyperref}
\begin{document}
\title{Nodeless superconductivity in noncentrosymmetric PbTaSe$_2$ single crystals}
\author{G. M. Pang}
\affiliation{Center for Correlated Matter and Department of Physics, Zhejiang University, Hangzhou 310058, China}
\author{M. Smidman}
\affiliation{Center for Correlated Matter and Department of Physics, Zhejiang University, Hangzhou 310058, China}
\author{L. X. Zhao}
\affiliation{Beijing National Laboratory for Condensed Matter Physics, Institute of Physics, Chinese Academy of Sciences, Beijing 100190, China}
\author{Y. F. Wang}
\affiliation{Center for Correlated Matter and Department of Physics, Zhejiang University, Hangzhou 310058, China}
\author{Z. F. Weng}
\affiliation{Center for Correlated Matter and Department of Physics, Zhejiang University, Hangzhou 310058, China}
\author{L. Q. Che}
\affiliation{Center for Correlated Matter and Department of Physics, Zhejiang University, Hangzhou 310058, China}
\author{Y. Chen}
\affiliation{Center for Correlated Matter and Department of Physics, Zhejiang University, Hangzhou 310058, China}
\author{X. Lu}
\affiliation{Center for Correlated Matter and Department of Physics, Zhejiang University, Hangzhou 310058, China}
\affiliation{Collaborative Innovation Center of Advanced Microstructures, Nanjing 210093, China}
\author{G. F. Chen}
\affiliation{Beijing National Laboratory for Condensed Matter Physics, Institute of Physics, Chinese Academy of Sciences, Beijing 100190, China}
\affiliation{Collaborative Innovation Center of Quantum Matter, Beijing 100084, China}
\author{H. Q. Yuan}
\email{hqyuan@zju.edu.cn}
\affiliation{Center for Correlated Matter and Department of Physics, Zhejiang University, Hangzhou 310058, China}
\affiliation{Collaborative Innovation Center of Advanced Microstructures, Nanjing 210093, China}


\begin{abstract}
We report an investigation of the superconducting order parameter of the noncentrosymmetric compound PbTaSe$_2$, which is believed to have a topologically nontrivial band structure. Precise measurements of the London penetration depth $\Delta\lambda(T)$ obtained using  a tunnel diode oscillator (TDO)  based method show an exponential temperature dependence at $T\ll T_c$, suggesting a nodeless superconducting gap structure. A single band $s$-wave model well describes the corresponding normalized superfluid density $\rho_s(T)$, with a gap magnitude of $\Delta(0)=1.85T_c$. This is very close to the value of $1.76T_c$ for weak-coupling BCS superconductors, indicating conventional fully-gapped superconductivity in PbTaSe$_2$.
\end{abstract}
\maketitle

Noncentrosymmetric superconductors (NCS), where the crystal structure does not have a center of inversion, display unique properties. In the presence of  antisymmetric spin-orbit coupling (ASOC), the pairing state is expected to be a mixture of spin singlet and triplet components, leading to unusual properties, even when the pairing has the conventional electron-phonon mechanism \cite{gor2001superconducting,frigeri2004superconductivity}. For instance, evidence for nodal superconductivity has been found in the weakly correlated NCS Li$_2$Pt$_3$B and Y$_2$C$_3$ \cite{yuan2006s,chen2011evidence}. Other NCS such as LaNiC$_2$ and BiPd, have been suggested to be fully gapped multiband superconductors, which although consistent with singlet-triplet mixing, this  may also arise in a purely singlet $s$-wave channel \cite{mondal2012andreev,JiaoBiPd,chen2013evidence}. Despite the anticipation of mixed parity pairing in NCS, BCS-like superconductivity is clearly observed in many of these materials, such as Re$_3$W, LaRhSi$_3$ and BaPtSi$_3$, where single band, fully-gapped behavior is reported \cite{Re3WBiswas,LaRhSi3Anand,bauer2009baptsi}.   The presence of singlet-triplet mixing has also been probed by looking for an anisotropic spin susceptibility via measurements of the upper critical field ($H_{c2}$) and Knight shift, as in the case of CeRhSi$_3$ and CeIrSi$_3$. \cite{CeRhSi3Hc2,CeIrSi3KS} In addition, NCS were also proposed to be good candidates for topological superconductivity \cite{NCSTop}, and there has been particular interest due to the possibility of gapless edge states and Majorana fermions. \cite{NCSMaj1,NCSMaj2}

Recently, the new NCS PbTaSe$_2$ has been the focus of much interest. The compound crystallizes in a noncentrosymmetric structure (space group $P\bar{6}m2$), containing hexagonal TaSe$_2$ layers \cite{TaSe2CDW}. The resistivity $\rho(T)$ of TaSe$_2$ shows a charge-density-wave (CDW) transition at around 122~K, which is gradually suppressed upon doping with Pb and pure PbTaSe$_2$ becomes superconducting below $T_c=3.7$~K \cite{ali2014noncentrosymmetric,sharafeev2015doping}. Meanwhile, the inversion symmetry of the crystal structure is also broken as a result of intercalation with Pb, making PbTaSe$_2$  a new NCS. Detailed comparisons between electronic structure calculations and angle resolved photoemission spectroscopy (ARPES) measurements reveal the presence of topological nodal lines in the band structure, near the Fermi level \cite{bian2015topological}. This is a result of the spin-orbit coupling lifting the degeneracy of intersecting bands, except along particular lines which are protected by the reflection symmetry of the crystal lattice, which gives rise to surface states observed in the ARPES measurements. In addition, ARPES measurements gave evidence for Dirac surface states and it was suggested that this may allow for the presence of Majorana bound states in the vortex cores of PbTaSe$_2$ \cite{chang2015topological}.

It is therefore of interest to determine whether there is any interplay between the topologically non-trivial band structure and the superconductivity. To find out whether this is a candidate for topological superconductivity, it is important to characterize the  superconducting order parameter of PbTaSe$_2$ through measurements of the bulk properties. In this article, we report the temperature dependence of the change of London penetration depth $\Delta\lambda(T)$ of high quality single crystal samples using a tunnel diode oscillator (TDO) based method. We find that exponential behavior is clearly observed at $T\ll T_c$, implying a nodeless superconducting energy gap. Further analysis reveals that both $\Delta\lambda(T)$ and the derived superfluid density $\rho_s(T)$ can be best fitted by a single band $s$-wave model, with a gap size close to the BCS value of $1.76T_c$, providing strong evidence for single band, weak-coupling nodeless superconductivity in the bulk of PbTaSe$_2$.

Single crystals of PbTaSe$_2$ were synthesized using a chemical vapor transport (CVT) process \cite{bian2015topological}. The temperature dependence of the electrical resistivity $\rho(T)$ was measured in a $^4$He system from 292~K to about 3~K, using a four-probe method. Measurements of the dc-magnetic susceptibility were performed in a superconducting quantum interference device (SQUID) magnetometer (MPMS-5T) in the temperature range of 2~-~6~K, with both field-cooling (FC) and zero-field-cooling (ZFC) in a small applied magnetic field of 10~Oe. 

The temperature dependence of the change in London penetration depth $\Delta\lambda(T)$ was precisely measured down to a base temperature of 0.35~K in a $^3$He cryostat using a TDO-based technique with an operating frequency of about 7~MHz and a noise level of about 0.1~Hz. Single crystal samples were cut into rectangular plates, with dimensions of $(500-900)\times(500-900)\times(5-10)~\mu$m$^3$. A tiny ac field of about 20~mOe was applied to the sample using the TDO coil, which is much smaller than the lower critical field $H_{c1}$, guaranteeing that the sample is entirely in the Meissner state and therefore the obtained $\Delta\lambda(T)$ can be regarded as the change of the London penetration depth. This quantity is  proportional to the TDO frequency change with  $\Delta\lambda(T)=G\Delta f(T)$, where the calibration constant $G$ is determined from the sample and coil geometry.

\begin{figure}[tb]
\begin{center}
  \includegraphics[width=\columnwidth]{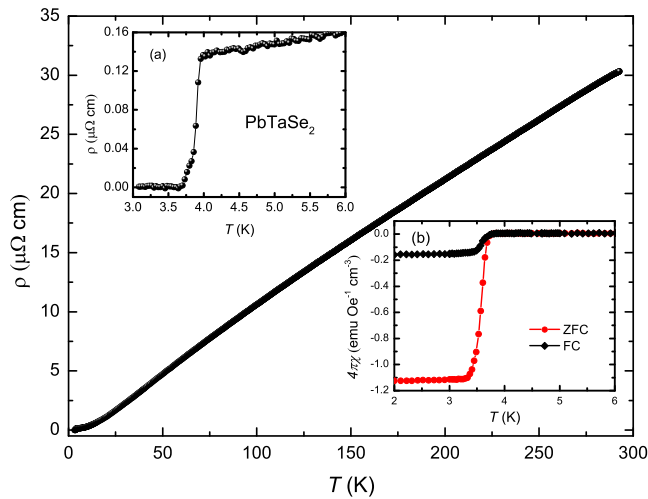}
\end{center}
	\caption{(Color online) The main panel shows the electrical resistivity $\rho(T)$  of PbTaSe$_2$ single crystals from room temperature down to around 3~K.  The insets show the low temperature behavior of (a) $\rho(T)$ and (b) magnetic susceptibility $4\pi\chi$, where sharp superconducting transitions are observed.}
   \label{figure1}
\end{figure}

Single crystals of PbTaSe$_2$ were characterized using both resistivity and magnetic susceptibility measurements, as shown in Fig.~\ref{figure1}. The temperature dependence of $\rho(T)$ displays metallic behavior, since it decreases upon cooling the sample from room temperature, with a very small residual resistivity of about $0.14~\mu\Omega$cm, just above the superconducting transition. As a result there is a large residual resistivity ratio (RRR) of $\rho(292 K)/\rho(4 K)$=220, which demonstrates the high quality of the single crystals. Upon following the method in Ref.~\onlinecite{orlando1979critical}, using a coherence length of $\xi=16.2$~nm and a normal state Sommerfeld coefficient $\gamma_n$=4.8~mJ/mol~K$^2$,\cite{cheng2015exotic} the small residual resistivity gives rise to a large mean free path of 10~$\mu$m. This is significantly  larger than the coherence length, indicating that the sample is in the clean limit. The presence of a superconducting transition is clearly displayed in the insets of Fig.~\ref{figure1}, onsetting at around 3.9~K in the resistivity and the magnetic susceptibility shows a clear diamagnetic signal, suggesting bulk superconductivity in PbTaSe$_2$. Here, the magnetic susceptibility slightly drops below the value of $-1$ for full diamagnetic shielding, which is likely due to demagnetization effects.

\begin{figure}[tb]
\begin{center}
  \includegraphics[width=\columnwidth]{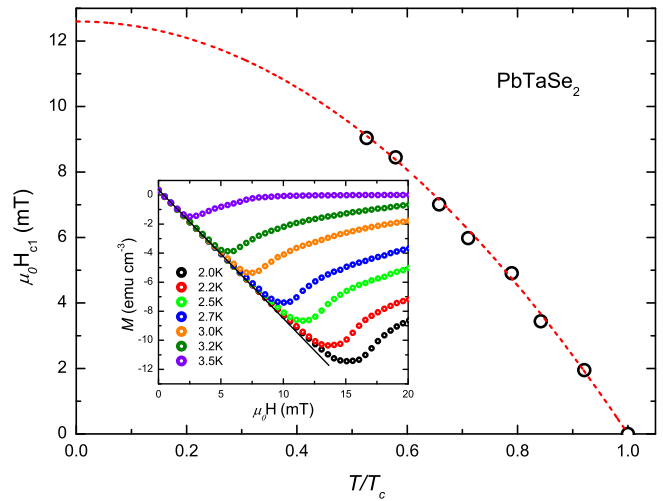}
\end{center}
	\caption{(Color online) Temperature dependence of the lower critical field $\mu_0H_{c1}$ of PbTaSe$_2$, defined as the field above which the field dependence of the magnetization $M(H)$ deviates from a linear decrease. The dashed red line shows a fit to the Ginzburg-Landau formula $H_{c1}(T)=H_{c1}(0)(1-(T/T_c)^2)$.  The inset shows $M(H)$ for fields applied in the $ab$ plane, at various temperatures in the superconducting state. }
   \label{figure2}
\end{figure}

In the inset of Fig.~\ref{figure2}, the field dependence of the magnetization is displayed, with the field applied in the $ab$ plane. In the case of type-${\mathrm{\uppercase\expandafter{\romannumeral2}}}$ superconductors, the magnetization $M(H)$ usually decreases linearly with increasing applied magnetic field for $\mu_0H<\mu_0$H$_{c1}$, where $\mu_0$H$_{c1}$ is the lower critical field, below which the sample is in the Meissner state. When $\mu_0$H$>\mu_0$H$_{c1}$, $M(H)$ deviates from linear behavior, due to the formation of vortices in the mixed state. We measured from zero field up to 20~mT at different temperatures in the superconducting state and $\mu_0$H$_{c1}$ was obtained from the field where this deviation occurs.

The temperature dependence of $\mu_0$H$_{c1}$ is shown in the main panel of Fig.~\ref{figure2}. The zero temperature lower critical field $\mu_0$H$_{c1}(0)=12.6$~mT was estimated from fitting H$_{c1}(T)$=H$_{c1}(0)(1-(T/T_c)^2)$, as shown by the red-dashed line in the figure. In Ginzburg-Landau theory, the magnetic penetration depth $\lambda$ is related to both the coherence length $\xi$ and lower critical field $\mu_0$H$_{c1}$, as described by $\mu_0\mathrm{H}_{c1}=\Phi/(4\pi\lambda^2)\ln{(\lambda/\xi+1/2)}$\cite{brandt2003properties}. Using the values of $\mu_0\mathrm{H}_{c1}$=12.6~mT and $\xi=16.2$~nm mentioned previously, $\lambda(0)$=198~nm was estimated and this value was used to calculate the superfluid density below. In addition, $\lambda(T)$ was estimated using the obtained values of $H_{c1}(T)$, where $\xi(T)$ was calculated from the reported values of $H_{c2}(T)$ \cite{cheng2015exotic}.

\begin{figure}[tb]
\begin{center}
  \includegraphics[width=\columnwidth]{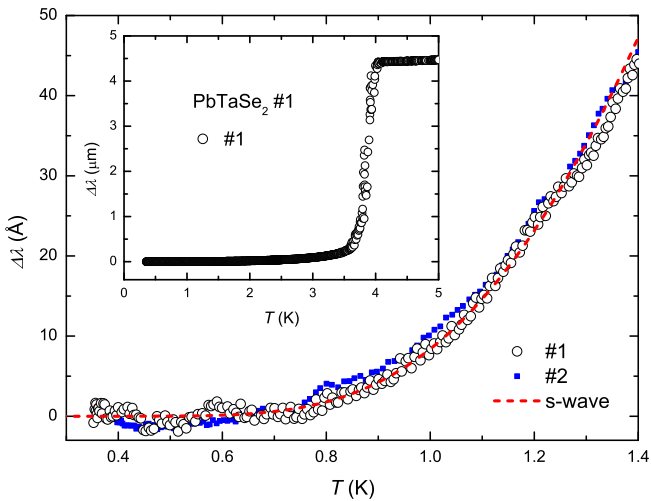}
\end{center}
	\caption{(Color online) The change of the London penetration depth $\Delta\lambda(T)$ as a function of temperature below $T_c/3$ for two samples of single crystal PbTaSe$_2$. The dashed line shows a fit to an $s$-wave model described in the text, which indicates conventional fully-gapped superconductivity. The inset shows $\Delta\lambda(T)$ from 5~K down to the base temperature for sample~$\#1$, which exhibits a sharp superconducting transition around 3.9~K, consistent with resistivity and magnetic susceptibility measurements.  }
   \label{figure3}
\end{figure}

The change of the London penetration depth as a function of temperature $\Delta\lambda(T)$ is shown for two samples in the inset of Fig.~\ref{figure3} from  5~K above $T_c$, down  to about 0.35~K, with calibration factors of $G~=~8.0$\AA/Hz and 10.0\AA/Hz for samples $\#1$ and $\#2$ respectively. A sharp decrease near 3.9~K indicates the onset of superconductivity, in agreement with the resistivity and susceptibility measurements and confirms the high sample quality. The main panel of Fig.~\ref{figure3} displays  the low temperature region below 1.4~K, where $\Delta\lambda(T)$ rapidly decreases upon cooling, before flattening below about 0.8~K, which suggests fully gapped superconductivity and the absence of low energy excitations. For an $s$-wave superconductor at $T\ll T_c$, $\lambda(T)$ can be approximated using
\begin{equation}
\Delta\lambda(T)=\lambda(T)-\lambda(0) = \lambda(0)\sqrt{\frac{\pi\Delta(0)}{2T}}\textrm{exp}(-\frac{\Delta(0)}{T}),
\label{equation1}
\end{equation}

\noindent where $\Delta(0)$ is the superconducting gap magnitude in units such that $k_B=1$\cite{prozorov2006magnetic}. The data for sample \#1 was fitted and the red dashed line in the figure shows that this expression gives excellent agreement with the experimental data below $T_c/3$, with a value of $\Delta(0)=1.71\mathrm{T_c}$, consistent with weakly coupled BCS superconductivity in PbTaSe$_2$.

\begin{figure}[tb]
\begin{center}
  \includegraphics[width=\columnwidth]{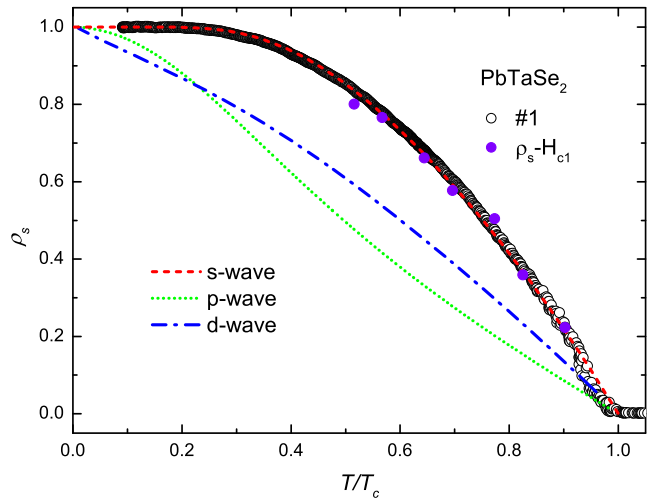}
\end{center}
	\caption{(Color online) Temperature dependence of the normalized superfluid density $\rho_s(T)$ for a single crystal of PbTaSe$_2$ (sample $\#1$), using a zero temperature value of the penetration depth of $\lambda(0)=198$~nm. The symbols represent the experimental data while the lines show the results of fitting with different models of the gap structure. The solid circles show  $\rho_s(T)$ obtained from the measurements of $H_{c1}(T)$, which are in a good agreement with the TDO data.}
   \label{figure4}
\end{figure}

To further characterize the superconducting gap symmetry of PbTaSe$_2$, we also analyzed the normalized superfluid density $\rho_s(T)$, which was converted from $\Delta\lambda(T)$ using $\rho_s=\lambda(0)^2/\lambda(T)^2$ and is displayed for sample \#1  in Fig.~\ref{figure4}. In addition the solid points show $\rho_s$ calculated from $\lambda(T)$ obtained from the values of $H_{c1}$, which are consistent with  $\rho_s$  obtained from the TDO measurements.  For a given gap function $\Delta_k$, $\rho_s$ can be calculated as:

\begin{equation}
\rho_{\rm s}(T) = 1 + 2 \left\langle\int_{\Delta_k}^{\infty}\frac{E{\rm d}E}{\sqrt{E^2-\Delta_k^2}}\frac{\partial f}{\partial E}\right\rangle_{\rm FS},
\label{equation2}
\end{equation}

\noindent where $f(E, T)=[1+{\rm exp}(E/T)]^{-1}$ is the Fermi distribution function and $\left\langle\ldots\right\rangle_{\rm FS}$ represents an average over the Fermi surface\cite{prozorov2006magnetic}. Here, the gap function is defined as $\Delta_k(T)=\Delta_0(T)g_k$, where $g_k$ contains the angular dependence and $g_k~=~1$, sin$\theta$ and cos$2\phi$ for the $s$-, $p$- and $d$-wave models respectively.    The temperature dependence of the gap [$\Delta_0(T)$]  is approximated as
\begin{equation}
\Delta_0(T)=\Delta_0(0){\rm tanh}\left\{1.82\left[1.018\left(T_c/T-1\right)\right]^{0.51}\right\},
\label{equation2}
\end{equation}

\noindent where $\Delta_0(0)$ represents the gap magnitude at zero temperature, which is the only adjustable parameter in the fitting\cite{carrington2003magnetic}.

The fitting results are also plotted in Fig.~\ref{figure4}. In the case of the $s$-wave model, the superconducting  gap is  isotropic. It is clearly shown that the experimental data is well fitted by this model, with a gap amplitude of $\Delta(0)=1.85\mathrm{T_c}$. This is consistent with the value obtained from fitting the penetration depth, and is comparable with other conventional BCS superconductors, suggesting that PbTaSe$_2$ is a fully-gapped, weakly coupled superconductor. On the other hand, a $p$-wave model with point nodes and a $d$-wave model with line nodes were also used for comparison, both of which predict $\rho_s$ to continue increasing with decreasing temperature due to existence of low energy excitations. These models strongly disagree with our results, giving further evidence for nodeless superconductivity in PbTaSe$_2$.

Our results indicate that the bulk superconductivity of  PbTaSe$_2$ is consistent with fully gapped, single band $s$-wave superconductivity with a gap size close to the weakly coupled BCS value. The nodeless bulk superconductivity is consistent with recent measurements  of the specific heat, which could also be fitted with exponential behavior at low temperatures \cite{raman2015single}. However, different strengths of the electron-phonon coupling were deduced, with strongly coupled behavior suggested from the large specific heat jump $\Delta C/\gamma T_c$ in Ref.~\onlinecite{raman2015single}, while the values in Refs.~\onlinecite{ali2014noncentrosymmetric,cheng2015exotic} are consistent with weak coupling, in line with our results. In addition, thermal conductivity measurements also indicate nodeless superconductivity \cite{wang2015multiple} and it was also suggested that the ``S"-shape in the field dependence of the thermal conductivity may indicate multi-band superconductivity in PbTaSe$_2$. However, from our measurements we find that the superfluid density can be well accounted for by a single gap model. 

The fact that these results are consistent with single band $s$-wave superconductivity would suggest that the pairing in noncentrosymmetric PbTaSe$_2$ is dominated by the spin singlet component. Indeed the lack of singlet-triplet mixing in PbTaSe$_2$ is somewhat of a puzzle, given the strong effect of the ASOC \cite{bian2015topological}. It should be noted that there are other NCS which are compatible with single band, BCS-like superconductivity and the reason for this behavior has yet to be resolved. It has previously been suggested that for NCS to have gapless edge states and Majorana modes, the triplet component should be larger than the $s$-wave singlet component \cite{NCSMaj1}. For $s$-wave superconductors in the presence of Rashba ASOC, it is necessary to apply a large magnetic field for Majorana fermions to be realized\cite{NCSMaj2}. Nevertheless, it was suggested in Ref.~\onlinecite{chang2015topological} that the topologically non-trivial nature of the electronic structure of PbTaSe$_2$ means that Majorana modes may be present in vortices, even if the bulk superconductivity is of an $s$-wave nature, as long as it is fully gapped. Given that our results support the existence of fully gapped superconductivity, further experiments are desirable to look for the presence of Majorana modes and other signatures of topological order.

To summarize, we have investigated the superconducting order parameter of the noncentrosymmertic superconductor PbTaSe$_2$. The change in the London penetration depth $\Delta\lambda(T)$ of single crystals was precisely measured down to 0.1$T_c$, which clearly shows an exponential temperature dependence below $T_c/3$, indicating fully gapped behavior. Both $\Delta\lambda(T)$ and its deduced superfluid density $\rho_s(T)$ can be described by a single band $s$-wave model, providing strong evidence for a nodeless superconducting gap structure in PbTaSe$_2$. A gap magnitude of $\Delta(0)=1.85\mathrm{T_c}$ is obtained from fitting the superfluid density, which slightly larger than the weak coupling BCS value of 1.76 and indicates the absence of strong electron-phonon coupling.

\begin{acknowledgments}
We thank C. Cao, S. H. Pan, D. L. Feng and D. F. Agterberg for interesting discussions. This work was supported by the National Basic Research Program of China (No.2011CBA00103), the National Natural Science Foundation of China (No.11474251 and No.11174245) and the Fundamental Research Funds for the Central Universities.

\end{acknowledgments}

\end{document}